\documentclass[12pt,thmsa]{article}
\usepackage{sw20aip}

\input tcilatex
\begin{document}

\title{Comment on''Propagator of two coupled general driven time-dependent
oscillators'' by C. F. Lo and Y.\ J.Wong}
\author{F. Benamira and L. Guechi \\
Laboratoire de Physique Theorique,\\
D\'epartement de Physique, Facult\'e des Sciences,\\
Universit\'e Mentouri, Route d'Ain El Bey, \\
Constantine, Alg\'eria.}
\maketitle

\begin{abstract}
We comment on several incorrect results given in a recent paper by Lo and
Wong. In particular, it is pointed out that their evaluation of the
propagator for two coupled general driven time-dependent oscillators is not
satisfactory. The correct expression can be obtained by applying an
appropriate time-dependent canonical transformation.

\end{abstract}

In a recent paper, Lo and Wong \cite{Lo} (hereafter referred to as LW) have
derived by means of a variant of the $su(1,1)$ Lie algebraic approach the
propagator associated to a system of two time-dependent coupled and driven
harmonic oscillators with time-varying angular frequencies and masses
described by the Hamiltonian

\begin{equation}
H(t)=\sum_{j=1}^{2}\left[ \frac{p_{j}^{2}}{2m_{j}(t)}+\frac{1}{2}%
m_{j}(t)\omega _{j}^{2}(t)x_{j}^{2}-m_{j}(t)f_{j}(t)x_{j}\right] +\lambda
(t)x_{1}x_{2}.  \label{a.1}
\end{equation}

In this comment, we would like to point out that the relation (3) in LW,
referred to as (LW 3) henceforth, is incorrect for time-dependent masses $%
m_j(t)$ and $\phi \neq $Const. Applying the time-dependent canonical
transformation (TDCT) defined by Eq. (LW 2), the new Hamiltonian has to be $%
\mathcal{H}(t)=H(t)+\frac \partial {\partial t}F_2,$ where $F_2$ is the
generating function of the canonical transformation. LW have omitted the
term $\frac \partial {\partial t}F_2$ in Eq. (LW 3). Therefore, Eqs (LW 5)
are not correct and the evaluation of the propagator is not satisfactory. In
particular, the Hamiltonian does not decompose into a sum of two uncoupled
driven oscillators, except in special cases.

A careful study of the dynamical system with the Hamiltonian (\ref{a.1}) has
been carried out in our paper \cite{bg} . For time-dependent masses the
optimal TDCT differs from the one used in LW and reads: 
\begin{equation}
\left\{ 
\begin{array}{c}
x_1=\frac{Q_1\cos \alpha (t)+Q_2\sin \alpha (t)}{\sqrt{m_1(t)}},\qquad x_2=%
\frac{-Q_1\sin \alpha (t)+Q_2\cos \alpha (t)}{\sqrt{m_2(t)}}, \\ 
p_1=\sqrt{m_1(t)}\left( P_1\cos \alpha (t)+P_2\sin \alpha (t)+\beta
_1(t)x_1\right) , \\ 
p_2=\sqrt{m_2(t)}\left( -P_1\sin \alpha (t)+P_2\cos \alpha (t)+\beta
_2(t)x_2\right) ,
\end{array}
\right.   \label{a.2}
\end{equation}
where the functions $\alpha (t),$ $\beta _1(t)$ and $\beta _2(t)$ can be
conveniently chosen to make separation of variables straightforward
possible. When $\alpha (t)\equiv {Const}$ and $\beta _j(t)=-\frac{\dot m_j(t)%
}{2\sqrt{m_j(t)}},\quad j=1,2,$ the new Hamiltonian takes the form: 
\begin{equation}
\mathcal{H}(t)=\sum_{j=1}^2\left[ \frac{P_j^2}2+\frac 12\Omega
_j^2(t)Q_j^2-F_j(t)Q_j\right] +\Gamma (t)Q_1Q_2,  \label{a.3}
\end{equation}
where the $\Omega _j$'s, $F_j$'s and $\Gamma $ are given by similar
relations as in LW but with the $\omega _j^2$'s replaced by $\widetilde{%
\omega }_j^2$'s as: 
\begin{equation}
\widetilde{\omega }_j^2(t)=\left[ \omega _j^2(t)+\frac 14\left( \frac{\dot
m_j^2(t)}{m_j^2(t)}-2\frac{\ddot m_j(t)}{m_j(t)}\right) \right] ;\,\quad
j=1,2.  \label{a.4}
\end{equation}

A comparaison of our relation (\ref{a.3}) with relation (LW 3) shows that
the canonical transformation can only be parametrized by a constant angle.
It is clear that the separation of variables in Eq. (\ref{a.3}) requires
that $\Gamma (t)\equiv 0$, i.e. $\lambda (t)=\frac 12\sqrt{m_1(t)m_2(t)}%
\left( \widetilde{\omega }_2^2(t)-\widetilde{\omega }_1^2(t)\right) \tan
\left( 2\alpha \right) ,$ which is a constraint on the parameters of the
system.

Following the path integral method described in Ref.\cite{bg} , the correct
expression of the propagator is given by 
\begin{eqnarray}
K(x_1^{\prime \prime },x_2^{\prime \prime },t^{\prime \prime };x_1^{\prime
},x_2^{\prime },t^{\prime })\!\!\! &=&\!\!\!\prod_{j=1}^2\sqrt{\frac{\left(
m_j^{\prime \prime }m_j^{\prime }\right) ^{\frac 12}}{2i\pi \hbar \rho
_j^{\prime \prime }\rho _j^{\prime }\sin \phi _j(t^{\prime \prime
},t^{\prime })}}\exp \left\{ -\frac i{4\hbar }\left. \frac{\dot m_j(t)}{%
m_j(t)}x_j^2\right| _{t^{\prime }}^{t^{\prime \prime }}\right\}  \nonumber \\
&&\!\!\!\!\!\!\!\!\!\!\!\!\times \exp \left\{ \frac i{2\hbar }\left( \frac{%
\dot \rho _j^{\prime \prime }}{\rho _j^{\prime \prime }}Q_j^{\prime \prime
2}-\frac{\dot \rho _j^{\prime }}{\rho _j^{\prime }}Q_j^{\prime 2}\right)
\right\}  \nonumber \\
&&\!\!\!\!\!\!\!\!\!\!\!\!\times \exp \left\{ \frac i{2\hbar \sin \phi
_j(t^{\prime \prime },t^{\prime })}\left[ \left( \frac{Q_j^{\prime \prime 2}%
}{\rho _j^{\prime \prime 2}}+\frac{Q_j^{\prime 2}}{\rho _j^{\prime 2}}%
\right) \cos \phi _j(t^{\prime \prime },t^{\prime })\right. \right. 
\nonumber \\
&&\!\!\!\!\!\!\!\!\!\!\!\!\!-\frac{2Q_j^{\prime \prime }Q_j^{\prime }}{\rho
_j^{\prime \prime }\rho _j^{\prime }}+2\frac{Q_j^{\prime \prime }}{\rho
_j^{\prime \prime }}\int_{t^{\prime }}^{t^{\prime \prime }}G_j(t)\sin \phi
_j(t,t^{\prime })dt  \nonumber \\
&&\!\!\!\!\!\!\!\!\!\!\!\!\!+2\frac{Q_j^{\prime }}{\rho _j^{\prime }}%
\int_{t^{\prime }}^{t^{\prime \prime }}G_j(t)\sin \phi _j(t^{\prime \prime
},t)dt  \nonumber \\
&&\!\!\!\!\!\!\!\!\!\!\!\!\!\!-\left. \left. 2\int_{t^{\prime }}^{t^{\prime
\prime }}\int_{t^{\prime }}^tG_j(t)G_j(\tau )\sin \phi _j(t^{\prime \prime
},t)\sin \phi _j(\tau ,t^{\prime })d\tau dt\right] \right\} ,  \label{a.5}
\end{eqnarray}
where $G_j(t)=F_j(t)\rho _j(t)$, $\phi _j(t",t^{\prime })=\int_{t^{\prime
}}^{t"}\frac{dt}{\rho _j^2}$ and $\rho _j(t)$ is the solution of the
auxiliary equation 
\begin{equation}
\stackrel{..}{\rho }_j+\Omega _j^2\rho _j=\frac 1{\rho _j^3}.
\end{equation}

\end{document}